\documentclass[aps,preprint,a4paper,onecolumn]{revtex4}%
\usepackage{amsfonts}
\usepackage[centertags]{amsmath}
\usepackage{amssymb}
\usepackage{graphicx}%
\begin{document}
\title{Using the mobile phone acceleration sensor in Physics experiments: free and
damped harmonic oscillations}
\author{Juan Carlos Castro-Palacio }
\affiliation{Departamento de F\'{\i}sica, Universidad de Pinar del R\'{\i}o. Mart\'{\i}
270, Esq. 27 de Noviembre, Pinar del R\'{\i}o, Cuba. CP-20100.}
\author{Luisberis Vel\'{a}zquez-Abad}
\affiliation{Departamento de F\'{\i}sica, Universidad Cat\'{o}lica del Norte. Av. Angamos
0610, Antofagasta, Chile.}
\author{Marcos H. Gim\'{e}nez}
\affiliation{Departamento de F\'{\i}sica Aplicada, Universitat Polit\`{e}cnica de
Val\`{e}ncia, Cam\'{\i} de Vera s/n, 46022, Val\`{e}ncia, Spain.}
\author{Juan A. Monsoriu}
\email{jmonsori@fis.upv.es}
\affiliation{Centro de Tecnolog\'{\i}as F\'{\i}sicas, Universitat Polit\`{e}cnica de
Val\`{e}ncia, Cam\'{\i} de Vera s/n, 46022, Val\`{e}ncia, Spain.}
\date{\today}

\begin{abstract}
The mobile acceleration sensor has been used to in Physics experiments on free
and damped oscillations. Results for the period, frequency, spring constant
and damping constant match very well to measurements obtained by other
methods. The Accelerometer Monitor application for Android has been used to
get the outputs of the sensor. Perspectives for the Physics laboratory have
also been discussed.

\end{abstract}
\maketitle

\section{Introduction}

Electronic\ portable or every day-use devices offer increasing perspectives
for the Physics laboratory at all teaching levels. This is the case of the
digital cameras \cite{monso}, web cams \cite{sham}, optical mouse of computers
\cite{rom,ng}, XBee transducers \cite{aya}, wiimote \cite{tom} and other game
console controllers \cite{van}. For instance, digital techniques have been
widely used to visualize Physics concepts \cite{vida1, vida2}. By using a
digital camera, a real process can be followed \cite{chung, grec}. Then, by
analyzing the recorded video a lot of information can be obtained such as
distances, time intervals and trajectories of objects.\ In this way, the study
and explanation of complex concepts to the students can be enormously
facilitated. More recently, many works related to the use of wireless devices
(such as the wiimote) in Physics teaching have been reported\ \cite{kaw, ocho,
tom}. The use of the wiimote for Physics experiments is based on its three
axis accelerometer which communicates with the game console using a bluetooth
device. Pioneer works about the use of accelerometers in the Physics
laboratory can be referred to Weltin \cite{wel} and Hunt \cite{hun1, hun2}.
The wiimote includes an infrared image sensor able to follow up to five
objects simultaneously, a bluetooth device and three accelerometers.

The wiimote gives Physics teachers a low cost way to track motion in a variety
of Physics experiments \cite{alex}, however, it is not a common device at the
Physics laboratories. In this paper, we study the possibility of using another
usual accelerometer which almost all students have with them, although they
may not realize it. We study free and damped oscillations by using the mobile
phone acceleration sensor and the air track. Mechanical oscillations is a very
important topic in most of undergraduate Physics courses. Many proposals of
Physics laboratory have been published in relation to this topic \cite{ono,
flo}. For their study, the air track is very useful since the friction force
can be appreciably decreased by creating a layer of air between the glider and
the track \cite{ber}. Even there are many works reporting experiments for the
study of oscillations, this is one the first ones that uses a mobile phone as
an accelerometer in Physics teaching.

The outline of the paper is the following. In section II, the experimental set
up is commented. It includes the introduction to the general features of the
free Acceleration Monitor mobile application. In section III, the free
harmonic oscillations are studied. This section is divided into two parts: the
first for the calculation of the period of the oscillations and the second for
the calculation of the spring constant. In section IV, the damping constant of
the damped oscillations is calculated. Finally, in section V, some conclusions
are drawn.

\section{Experimental set up}

A photograph of the experimental set up is shown in figure 1. The mobile is
placed on the glider of the air track and connected to a fixed end by a
spring. Once the air flow goes through the air track the friction is
considerably decreased. Under this situation, the glider with the mobile on
can oscillate almost freely upon perturbation of the system. The mobile that
has been used for the measurements is a Smart phone, LG-E510 bearing an
Android version $2.3.4$. The mobile mass is $124.0$ $g$ and the air track
glider mass is $180.6$ $g$. The mass of the glider can be changed by adding
weights at both sides. In this way, the frequency of the free oscillations is
changed. \ \ 

For the interaction with the mobile sensor, the free Android application
"Accelerometer Monitor ver $1.5.0$" has been used. This application takes
$348$ $kB$ of SD card memory and it can be downloaded from Google play website
at ref. \cite{ac-mon}. This application reports the vibrations of the mobile
phone in real time by registering the acceleration components on $x,y$ and $z$
- axis at each time step. The component of the acceleration of gravity can be
removed from the data. The precision in the measurement of the acceleration is
$\delta a=0.01197$ $m/s^{2}$ and for the time is $\delta t=0.02$ $s$. Since
the oscillations take place along the $y$-axis, the values of the acceleration
for the $x$ and $z$-axis remain very close to zero. This application also
allows saving the output data to a file, from which the data can used for
further analysis.

Once the application has been downloaded to the mobile device, a small test
can be done to check its well functioning. If the mobile is left quiet on a
horizontal surface, the application output curves for the acceleration should
indicate values very close to zero for all axis.

For the case of damped oscillations, some dissipation of the amplitude of the
oscillations can be obtained by decreasing the air flow at the air track.

\section{Free harmonic oscillations}

The experimental results of the acceleration $a(t)$ obtained with the
Accelerometer Monitor mobile application have been fitted to the following equation:%

\begin{equation}
a(t)=A\sin(\omega_{0}t+\phi_{0}) \label{ec1}%
\end{equation}
where $A$ is the acceleration amplitude, $\omega_{0}$ is the angular frequency
and $\phi_{0}$ the phase constant. The parameters have been obtained by using
a least square fitting. In figure 2, the screen of the Accelerometer Monitor
application while displaying free harmonic oscillations has been shown.

In order to start the oscillations in the system, the glider with the mobile
was shifted to the right and then left free. The six different masses were
obtained by adding weights to both sides of the glider. The output of the
mobile application with the acceleration data is collected in an ASCII\ file.
First, a heading with some information such as "saving start date and time",
"sensor resolution" and "sensor vendor", can be found. Following the heading,
the sensor measurements are shown. The first three columns indicate the
acceleration in the three perpendicular axis, $x$ (perpendicular to the
device, positive to the right), $y$ (along the device, positive upward) and
$z$ (perpendicular to $x$ and $y$ axis, with positive direction as coming out
perpendicularly from the device display). In figure 3, a fragment of the
output file is shown.

The scatter of the data points and the fit curve are shown in figure 4. In
table I, the parameters and their errors from the fitting to equation
\ref{ec1} are registered. The quality of the fit can be seen in the values of
the regression coefficient $R^{2}$ whose values are close to $1$.

\begin{table}[ptb]
\caption{Parameters and their errors from the fitting of the acceleration data
for free oscillations.}%
\label{Table 1}
\[%
\begin{tabular}
[c]{cccccc}\hline\hline
& $(m\pm0.0001)(kg)$ & $(A\pm\delta A)$ $m/s^{2}$ & $(\omega_{0}\pm
\delta\omega_{0})$ $rad/s$ & $(\phi_{0}\pm\delta\phi_{0})$ $rad$ & $R^{2}%
$\\\hline
$m_{1}$ & $0.3045$ & $1,082\pm0,008$ & $24,747\pm0,010$ & $-0,797\pm0,015$ &
$0,9938$\\
$m_{2}$ & $0.4043$ & $1,204\pm0,006$ & $21,546\pm0,005$ & $2,479\pm0,009$ &
$0,9968$\\
$m_{3}$ & $0.5004$ & $1,598\pm0,006$ & $19,408\pm0,004$ & $-0,377\pm0,007$ &
$0,9984$\\
$m_{4}$ & $0.6084$ & $1,171\pm0,007$ & $17,669\pm0,006$ & $-0,397\pm0,011$ &
$0,9953$\\
$m_{5}$ & $0.6285$ & $0,856\pm0,007$ & $17,371\pm0,006$ & $2,553\pm0,016$ &
$0,9875$\\
$m_{6}$ & $0.6961$ & $1,542\pm0,008$ & $16,526\pm0,007$ & $2,830\pm0,011$ &
$0,9966$\\\hline\hline
\end{tabular}
\
\]
\end{table}

\subsection{Calculation of the period $T$}

In table II, the calculation of the period by using the fitted frequency and
the period obtained directed from the photometer are shown. Results indicate a
very good agreement between the values since discrepancies are less than $1$
$\%$ in most of cases.

\begin{table}[ptb]
\caption{Periods calculated from the fit (column 2) and the period obtained
with the photometer (column 3) are shown. In the fourth column, the percentage
discrepancy has been included.}%
\label{Table 2}
\[%
\begin{tabular}
[c]{cccc}\hline\hline
& $(T_{fit}\pm\delta T_{fit})$ $s$ & $(T_{Photo}\pm0.001)$ $s$ &
$D(\%)$\\\hline
$m_{1}$ & $0,2539\pm0,0001$ & $0,259$ & $1,99$\\
$m_{2}$ & $0,2916\pm0,0001$ & $0,291$ & $0,21$\\
$m_{3}$ & $0,3237\pm0,0001$ & $0,323$ & $0,23$\\
$m_{4}$ & $0,3556\pm0,0001$ & $0,356$ & $0,11$\\
$m_{5}$ & $0,3617\pm0,0001$ & $0,365$ & $0,91$\\
$m_{6}$ & $0,3802\pm0,0002$ & $0,380$ & $0,05$\\\hline\hline
\end{tabular}
\ \
\]
\end{table}

\subsection{Calculation of the spring constant $k$}

Once the values of the mass ($m$) and the frequency of the free oscillations
($\omega_{0}$) have been obtained, the spring constant $k_{fit}$ can be
calculated. For that purpose, a least square linear regression to the equation
$\omega_{0}^{2}=k_{fit}m^{-1}$ has been performed using the values of table
III. The string constant has also been calculated by hanging a mass ($m$) from
the spring and measuring the shift in the position ($x$). For the values,
$m=\left(  500,2\pm0.1\right)  $ $g$ and $x=\left(  2,6\pm0,1\right)  $ $cm$,
we obtained $k=mg/x=\left(  189\pm7\right)  $ $N/m$. The results for the
spring constant $k$ and their errors have been shown in table IV.

\begin{table}[ptb]
\caption{Data for the calculation of the spring constant.}%
\label{Table 3}
\[%
\begin{tabular}
[c]{ccc}\hline\hline
& $(m^{-1}\pm\delta(m^{-1}))$ $kg^{-1}$ & $\ (\omega_{0}^{2}\pm\delta
\omega_{0}^{2})$ $rad^{2}/s^{2}$\\\hline
$m_{1}$ & $3,2841\pm0,0011$ & $612,4\pm0,5$\\
$m_{2}$ & $2,4734\pm0,0006$ & $464,2\pm0,2$\\
$m_{3}$ & $1,9984\pm0,0004$ & $376,7\pm0,1$\\
$m_{4}$ & $1,6437\pm0,0003$ & $312,2\pm0,2$\\
$m_{5}$ & $1,5911\pm0,0003$ & $301,8\pm0,2$\\
$m_{6}$ & $1,4366\pm0,0002$ & $273,1\pm0,2$\\\hline\hline
\end{tabular}
\]
\end{table}

\begin{table}[ptb]
\caption{Results for the spring constant, from the fit (column 1) and using a
weight (column 2). In the third column, the percentage discrepancy between
both values has been shown.}%
\label{Table 4}
\[%
\begin{tabular}
[c]{ccc}\hline\hline
From the fit & Using a weight & \\\hline
$(k_{fit}\pm\delta k_{fit})$ $N/m$ & $(k\pm\delta k)$ $N/m$ & $D(\%)$\\
$187.9\pm0.6$ & $189\pm7$ & $0,58$\\\hline\hline
\end{tabular}
\]
\end{table}

\section{Damped harmonic oscillations}

The damped harmonic oscillations of the acceleration $a(t)$ can be represented
by the following equation:%

\begin{equation}
a(t)=De^{-\gamma t}\sin(\omega t+\phi) \label{ec3}%
\end{equation}
where $D\ $is the acceleration amplitude for $t=0$ $s$, $\gamma$ is the
damping constant and $\phi$, the phase.

In figure 5, the screen of the Accelerometer Monitor application while
displaying damped harmonic oscillations has been shown. In figure 6, the data
points and the fit curve are illustrated. In the table V, the parameters and
their errors from the fitting to equation \ref{ec3} using the mass $m_{6\text{
}}$ are registered. The value of $R^{2}$ which indicates the quality of the
fitting has also been included.

\begin{table}[ptb]
\caption{Parameters and their errors from the fitting of the acceleration data
for damped oscillations.}%
\label{Table 5}
\[%
\begin{tabular}
[c]{cccccc}\hline\hline
& $(D\pm\delta D)$ $m/s^{2}$ & $(\gamma\pm\delta\gamma)$ $s^{-1}$ &
$(\omega\pm\delta\omega)$ $rad/s$ & $(\phi\pm\delta\phi)$ $rad$ & $R^{2}%
$\\\hline
$m_{6}$ & $1,64\pm0,02$ & $0,58\pm0,02$ & $16,52\pm0,01$ & $-0,56\pm0,01$ &
$0,9816$\\\hline\hline
\end{tabular}
\]
\end{table}

\begin{table}[ptb]
\caption{Results for the relaxation times. The percentage discrepancy between
both values has also been included.}%
\[%
\begin{tabular}
[c]{ccc}\hline\hline
$(\tau_{fit}\pm\delta\tau_{fit})$ $s^{-1}$ & $(\tau_{Da}\pm\delta\tau_{Da})$
$s^{-1}$ & $D(\%)$\\\hline
$1,71\pm0.02$ & $1,70\pm0,04$ & $0,59$\\\hline\hline
\end{tabular}
\]
\end{table}

The relaxation time, $\tau$, is the inverse of the damping constant,
$\tau=1/\gamma$. It can also be derived from the following equation:%

\begin{equation}
\tau=\frac{T_{Da}T_{fit}}{2\pi}\frac{1}{(T_{Da}^{2}-T_{fit}^{2})^{1/2}}
\label{ec4}%
\end{equation}
where, $T_{fit}$, is the period of the free oscillations (see table II) and
$T_{Da}=2\pi/\omega$, the period of the damped oscillations. In table VI,~the
relaxation times, obtained from the fitting ($\tau_{fit}=1/\gamma$) and from
equation \ref{ec4} ($\tau_{Da})$, are shown. Results indicate a very good agreement.

\section{Conclusions}

Free and damped oscillations have been studied in a very simple way by using
the mobile phone acceleration sensor. For this purpose, the free Accelerometer
Monitor application for Android operating system have been used. Results for
the period, frequency and the spring constant\ using the mobile sensor data
show a very good agreement with the classical way of measuring this quantities
in the Physics laboratory. For instance, the period obtained from the fitting
is compared to the direct measurement of the photometer and the spring
constant, calculated from the fit, is compared to the spring constant
calculated from a gravity method using a weight. On the other hand, the
damping constant as obtained from the fitting of the mobile phone data allows
corroborating theoretical expressions on damped oscillations. This fact offers
very good perspectives for the use of the acceleration mobile phone sensor in
the General Physics laboratory. For example, the mobile phone as accelerometer
sensor can be used to study the most common Physics laboratory pendulums such
as the simple, physical and torsion pendulums. It can also be used for the
study of two dimensional harmonic oscillations, for example, with the use of
the air table \cite{bob}.

\begin{acknowledgments}
Authors would like to thank the Institute of Education Sciences, Universitat
Polit\`{e}cnica de Val\`{e}ncia (Spain), for the support of the Teaching
Innovation Group, MoMa.
\end{acknowledgments}

\section{Figure captions}

Fig.1. Photograph of the experimental set up used in the experiments. In the
photograph (1) is the smart phone, (2) the glider, (3) the air track, (4) the
string, (5) the photometer, and (6) the fixed end.

Fig. 2. Photograph of the experimental set up used in the experiments. In the
photograph (1) is the smart phone, (2) the glider, (3) the air track, (4) the
string, (5) the photometer, and (6) the fixed end.

Fig. 3. Fragment of the output data file of the Accelerometer Monitor mobile
phone application.

Fig. 4. As an example of the data points (scatter) and fit curves (solid line)
of \ the acceleration \textit{versus} time, the case of the mass 3 has been shown.

Fig. 5. Photograph of the smart phone displaying a view of the Acceleration
Monitor application for the case of damped harmonic oscillations.

Fig. 6. Data points of the acceleration (scatter) \textit{versus }time for
damped oscillations. The solid line represent the fit curve.

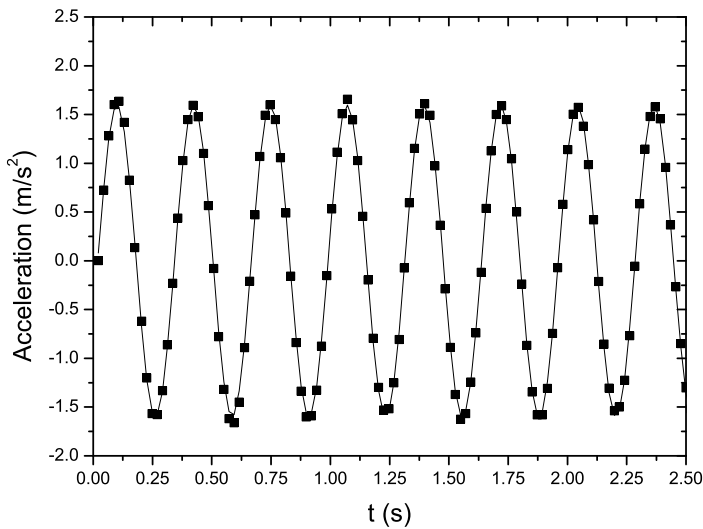
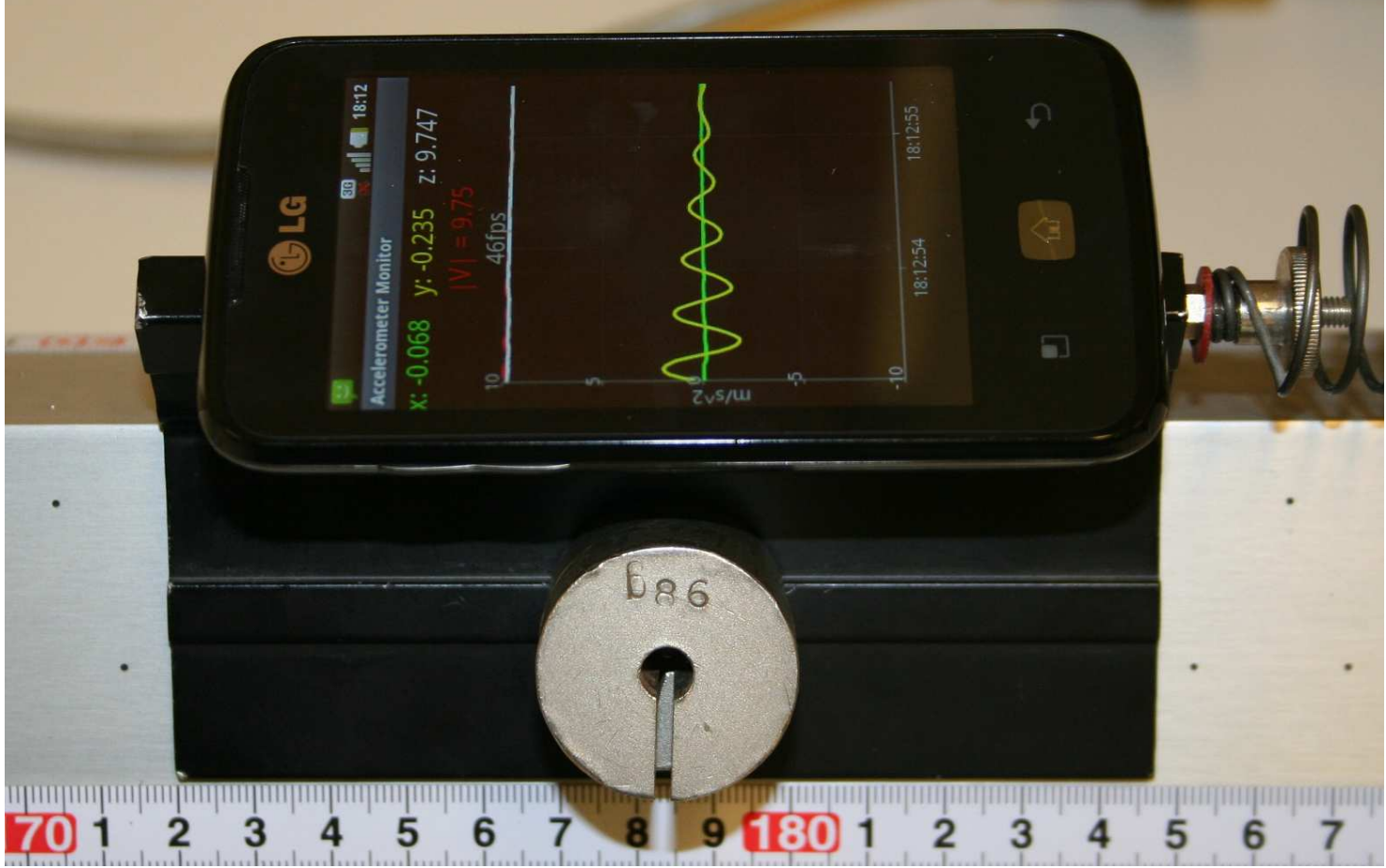
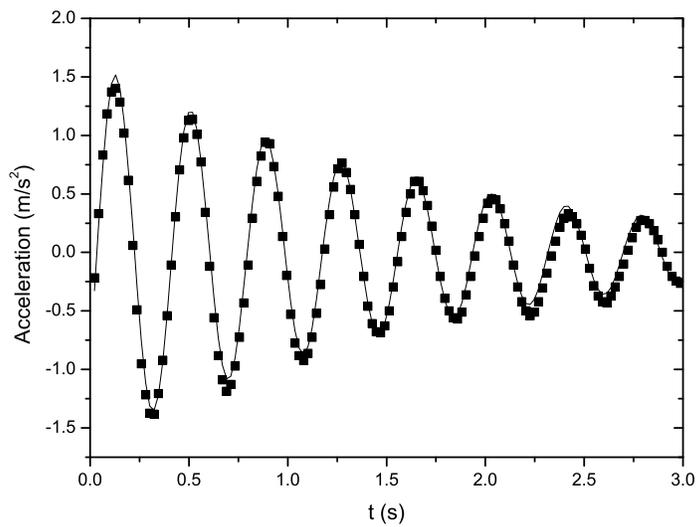
\end{document}